# ChestX-Det10: Chest X-ray Dataset on Detection of Thoracic Abnormalities

Jingyu Liu, Jie Lian, Yizhou Yu

*Deepwise AI Lab*

Instance level detection of thoracic diseases or abnormalities are crucial for automatic diagnosis in chest X-ray images. Most existing works on chest X-rays focus on disease classification and weakly supervised localization. In order to push forward the research on disease classification and localization on chest X-rays. We provide a new benchmark called ChestX-Det10, including box-level annotations of 10 categories of disease/abnormality of ∼ 3,500 images. The annotations are located at https://github.com/Deepwise-AILab/ChestX-Det10-Dataset.

## 1  Introduction

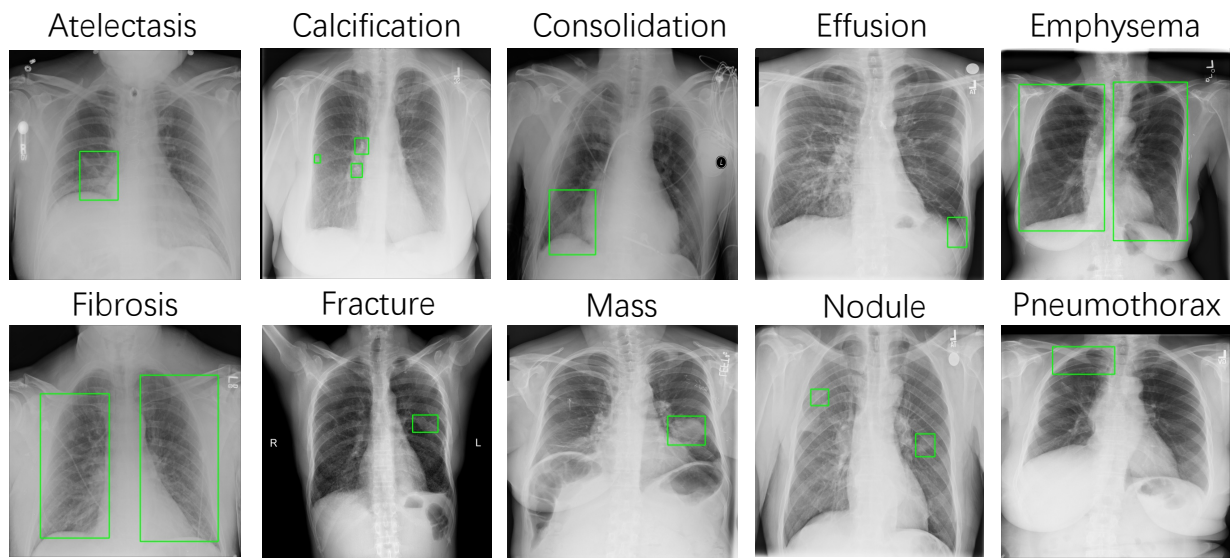

Figure 1: Examples of annotated diseases of 10 categories in ChestX-Det10.

ChestX-Det10 is a subset with box annotations of NIH ChestX-14 [1]. We select 3,543 images from NIH ChestX-14 and invite three board-certified radiologists to annotate them with 10 common categories of diseases or abnormalities. All data are selected to cover the appearance range as wide as possible. Figure 1 shows examples of the annotated 10 categories.

## 2 Annotation Procedure and Dataset Statistics

For annotation, we split three board certified radiologists into two roles.

**Commitee: composed of two radiologists** Each chest X-ray image is annotated by both of them individually. The two radiologists are mutually blind to the annotation.

**Judge: the most experienced radiologist** The judge can select annotations from above, and can add annotation if there is more. We call the manual delineations generated in this way the 'gold standard' for both training and testing purposes, while being mindful of the caveat that there are potential subjective variants within the annotation.

| Datesets | Sets | Positive | Negative | Total |
|---|---|---|---|---|
| ChestX-Det10 | Training Set | 2320 | 681 | 3001 |
| | Test Set | 459 | 83 | 542 |

Table 1: Positive/Negative samples statistics.

ChestX-Det10 includes samples in different angles, whiteness, and scanning conditions. We randomly split the data into 3001 images for training and 542 images for testing. We provide more detailed dataset statistics in Table 1 and Table 2. Table 1 shows number of positive and negative images in ChestX-Det10, where positive/negative images are w/o labeled diseases. Table 2 shows instance number of each disease. Please be noted that multiple diseases might share the same region of box.

| | Sets | Atelectasis | Calcification | Consolidation | Effusion | Emphysema |
|---|---|---|---|---|---|---|
| ChestX-Det10 | Training Set | 289 | 280 | 2091 | 1720 | 232 |
| | Test Set | 51 | 67 | 446 | 372 | 66 |
| | Sets | Fibrosis | Fracture | Mass | Nodule | Pneumothorax |
| | Training Set | 618 | 546 | 129 | 789 | 169 |
| | Test Set | 120 | 115 | 31 | 166 | 42 |

Table 2: Abnormalities statistics.

## 3  Experiments

In this section, we use the Faster R-CNN with FPN [2] as the baseline and report the detection results on ChestX-Det10. The backbone ResNet-50 is pretrained on ImageNet [3]. The experiment is implemented using Pytorch on 4 TITAN-V GPUs.

For training, we apply stochastic gradient descent (SGD) with a weight decay of 0.0001 and momentum of 0.9 to optimize the model. The first conv layers of FPN is frozen. We train 50 epochs with image batch-size of 2 on each GPU. The learning rate starts at 0.01, and reduce by a factor of 10 after 20 and 40 epochs. During training, we adopt multi-scale sampling (shorter side=800 $\sim$ 1400) for all images. At testing stage, the shorter side of image is fixed at 1200.

**Evaluation:** We borrow the evaluation metric of AP50 (IOU=0.5) used in general object detection. Table 3 shows detection results at AP50 of each category.

| | % | Atelectasis | Calcification | Consolidation | Effusion | Emphysema | Fibrosis |
|---|---|---|---|---|---|---|---|
| Baseline | | 30.6 | 42.0 | 53.7 | 44.0 | 66.2 | 33.4 |
| | | Fracture | Mass | Nodule | Pneumothorax | **Mean** | |
| | | 38.8 | 43.4 | 23.3 | 26.1 | **40.2** | |

Table 3: Detection results at AP50 of each category on ChestX-Det10. The baseline is Faster R-CNN with FPN.

For more practical usage, we also provide recall (sensitivity) at fixed false positives per image for each category in Table 4. In particular, we set the rate of instance-level FP/image to be 0.1 at IOU50.

| | Atelectasis | Calcification | Consolidation | Effusion | Emphysema | Fibrosis |
|---|---|---|---|---|---|---|
| Baseline | 0.431 | 0.597 | 0.371 | 0.277 | 0.758 | 0.342 |
| | Fracture | Mass | Nodule | Pneumothorax | **Mean** | |
| | 0.426 | 0.548 | 0.301 | 0.310 | **0.436** | |

Table 4: Recall@0.1fp/image at AP50 of each category on ChestX-Det10. The baseline is Faster R-CNN with FPN.

Moreover, we report the average prediction time per image on 542 samples: Using Faster R-CNN with FPN as the baseline (single GPU, single scale, batch=1), time is 0.101s/image.

# 4  Conclusion

Our ChestX-Det10 is the first chest X-Ray dataset with instance-level annotations. Using a fully supervised method to improve chest X-rays detection performance to a level that is more useful in the clinical environment. We believe the dataset can push forward the research on thoracic diseases detection on chest X-rays.

# References

[1] Xiaosong Wang, Yifan Peng, Le Lu, Zhiyong Lu, Mohammadhadi Bagheri, and Ronald M Summers. Chestx-ray8: Hospital-scale chest x-ray database and benchmarks on weakly-supervised classification and localization of common thorax diseases. In *Proceedings of the IEEE Conference on Computer Vision and Pattern Recognition*, pages 2097–2106, 2017.

[2] T.-Y. Lin, P. Dollár, R. Girshick, K. He, B. Hariharan, and S. Belongie. Feature pyramid networks for object detection. *In CVPR*, 2017.

[3] O. Russakovsky, J. Deng, H. Su, J. Krause, S. Satheesh, S. Ma, Z. Huang, A. Karpathy, A. Khosla, and M. Bernstein. Imagenet large scale visual recognition challenge. *International Journal of Computer Vision*, 115(3):211–252, 2015.